# Strain-dependent structural and electronic reconstructions in long-wavelength WS$_2$ moiré superlattices

Kai-Hui Li[1,2,§], Fei-Ping Xiao[3,§], Wen Guan[1,2], Yu-Long Xiao[1,2], Chang Xu[1,2], Jin-Ding Zhang[1,2], Chen-Fang Lin[1,2], Dong Li[1,2], Qing-Jun Tong[3,†], Si-Yu Li[1,2,†], An-Lian Pan[1,2,†]

[1]Key Laboratory for Micro-Nano Physics and Technology of Hunan Province, College of Materials Science and Engineering, Hunan University, Changsha 410082, People's Republic of China

[2]Hunan Institute of Optoelectronic Integration, Hunan University, Changsha 410082, People's Republic of China

[3]School of Physics and Electronics, Hunan University, Changsha 410082, People's Republic of China

[§]These authors contributed equally to this work.

[†]Correspondence and requests for materials should be addressed to Q.J.T. (e-mail: tongqj@hnu.edu.cn), S.Y.L (e-mail: lisiyu@hnu.edu.cn) and A.L.P. (e-mail: anlian.pan@hnu.edu.cn).

**In long-wavelength moiré superlattices of stacked transition metal dichalcogenides (TMDs), structural reconstruction ubiquitously occurs, which has reported to impact significantly their electronic properties. However, complete microscopic understandings of the interplay between the lattice reconstruction and alteration of electronic properties, and their further response to external perturbations in the reconstructed TMDs moiré superlattice are still lacking. Here, using scanning tunneling microscopy (STM) and scanning tunneling spectroscopy (STS) combined with first-principles calculation, we study the strain-dependent structural reconstruction and its correlated electronic reconstruction in long-wavelength H-type WS$_2$ moiré superlattice at nanometer scale. We observe that the long-wavelength WS$_2$ moiré superlattices experiencing strong atomic reconstruction transform into a hexagonal array of screw dislocations separating large-sized H-stacked domains. Both the geometry and the moiré wavelength of the moiré superlattice are dramatically tuned by external intralayer heterostrain in our experiment. Remarkably, the STS measurements further demonstrate that the location of the *K* point in conduction band is modulated sensitively by strain-induced lattice deformation at nanometer scale in this system, with the maximum energy shift reaching up to 300 meV. Our results highlight that intralayer strain plays a vital role in determining structural and electronic properties in TMD moiré superlattice.**

Moiré superlattices in transition metal dichalcogenides (TMDs) hetero-bilayers and twisted homo-bilayers have been regarded as an excellent model system to study and manipulate intriguing quantum phenomena[1,2], such as moiré excitons[3-7] and strongly correlated electron states[8-18]. The key physics behind these novel quantum states is the alteration of the electronic properties by atomic moiré structures, which can be further engineered by twist angle and strain[19-27]. In marginally twisted TMDs bilayers, the long-wavelength moiré superlattices are demonstrated to experience strong structural reconstruction into large-sized commensurate stacking domains divided by a dislocation network[28-34], due to the interplay between interlayer van der Waals (vdW) interaction and intralayer lattice deformation[35-37]. The reconstruction in these marginally twisted TMDs bilayers not only reshapes greatly the moiré structure, but also changes significantly the electronic structure, leading to qualitatively different physical behaviors from the unreconstructed moiré superlattices[20,28-31]. The intralayer strain is believed to play a central role on the structure reconstruction of long-wavelength moiré superlattice, however, a deep microscopic exploration of the strain-dependent atomic reconstruction and its correlation with electronic modulation is still lacking.

In this letter, by using scanning tunneling microscopy (STM) combined with first-principles calculations, we study the lattice reconstruction and the resulting electronic modulation of long-wavelength H-type $WS_2$ moiré superlattices, in the present of external intralayer heterostrain at nanometer scale. In our experiment, the H-type $WS_2$ moiré superlattices experience strong structural reconstruction to form large-sized H-stacked domains separated by a hexagonal array of screw dislocations. The existence of external intralayer heterostrain deforms the topography of the reconstructed moiré superlattice and tunes the moiré wavelength continuously from tens of nanometers to hundreds of nanometers. Our STS measurements further demonstrate that the energy location of $K$ point in conduction band is sensitively modulated by strain-induced lattice deformation, with the maximum energy modulation reaching up to 300meV in the 1D dislocation of the sharply deformed moiré superlattice. Our result paves the way to engineer exotic structural and electronic properties by intralayer strain in TMDs moiré superlattices.

Our STM experiments were carried out on an H-stacked multilayer $WS_2$ sheet which is mechanically exfoliated from an H-phase bulk $WS_2$ crystal and transferred onto an Au deposited $SiO_2/Si$ substrate covered with a thin Cr adhesion layer at $T$ = 78 K (see

Methods in the Supplementary Material). The surface layer of the multilayer $WS_2$ sheet is disturbed during exfoliation and transfer progresses, and generates misalignment relative to the underlying layers. Fig. 1(a) is a large-scaled (700 nm×400 nm) STM image detected on our sample, showing an interesting topography that contains various long-wavelength moiré superlattices with domain geometry transforming from nearly regular hexagon (left part) to sharply deformed hexagon (right part). Since the hexagonal commensurate domain is the unique character in reconstructed moiré superlattice of marginally twisted 2H bilayer TMDs[28,29], we identify that there is a marginal twist angle between the misaligned topmost layer and the underlying multilayer $WS_2$ sheet in our sample. The formation of sharply deformed moiré superlattice in the right part of Fig. 1(a) indicates the existence of inhomogeneous intralayer strain in the sample, which tunes the moiré wavelength continuously from about 30nm to hundreds of nanometers as shown in Fig. 1(a) and Fig. S1 of the Supplementary Material.

We firstly study the atomic reconstruction in the long-wavelength H-type $WS_2$ moiré superlattice. Fig. 1(b) shows a magnified STM image of the reconstructed moiré superlattice containing nearly regular hexagonal domains, with the moiré wavelengths of $L_1 = 31.60$nm, $L_2 = 31.22$nm and $L_3 = 38.67$nm in three different zigzag directions, respectively. The small anisotropy in the moiré wavelength indicates the existence of a small external intralayer heterostrain in this region[38,39]. The twist angle can be approximately estimated as $\theta \approx 0.57°$ according to $L = a/[2\sin(\theta/2)]$[26,40-43], where $a \approx 0.316$nm is the lattice constant of $WS_2$. Due to the strong structural reconstruction, large-sized hexagonal H-stacked domains are formed and separated by a network of dislocations corresponding to one-dimensional domain walls (1D DWs) [Fig. 1(d)]. The nodes of these 1D DWs are the seeds of the XX′-stacked region (the configuration that $S$ atoms lie on top of each other) and the MM′-stacked region (the configuration that W atoms lie on top of each other), respectively, as shown in the atomic-resolved Figs. 1(e), 1(f). The difference in domain sizes of these three high-symmetry stackings in the reconstructed moiré superlattice implies that H stacking has the lowest formation energy, while MM′ stacking is less energetically favorable and XX′ stacking is most energetically unfavorable[23,28,32]. As for the sharply deformed moiré superlattices in the right part of Fig. 1(a), one of the 1D DWs is dramatically shortened in length with XX′- and MM′-stacked regions almost overlapping, indicating the existence of a strong external heterostrain in these regions.

To further understand this interesting moiré topography in our experiment, we use first-principles calculations to simulate the structural evolution of moiré superlattice in marginally twisted bilayer WS$_2$ ($\theta$ = 0.57°) considering structural reconstruction and external intralayer heterostrain (See Methods). In the absence of lattice reconstruction and external heterostrain in Fig. 1(g), the marginally twisted WS$_2$ bilayer has smoothly varying rigid-lattice moiré superlattices with the domain sizes of 2H, XX′, MM′ stackings being exactly same. The rigid-lattice moiré superlattices are usually observed in large-angle twisted 2D materials with small moiré wavelength[26,40-44]. Once lattice reconstruction is considered, the 2H-stacked region dramatically expands in size and becomes a hexagonal domain in Fig. 1(h), while the XX′-stacked region shrinks as a minimum domain and the MM′-stacked region has an intermediate size. When adding a 0.4% intralayer compression heterostrain along zigzag direction, the moiré structure deforms slightly with its wavelengths being anisotropic in three zigzag directions [Fig. 1(i)], which is accorded with our experimental STM result in Fig. 1(b). Further increasing the compression heterostrain to 0.9%, the moiré deformation becomes more drastic until XX′-stacked region and MM′-stacked region almost overlap in Fig. 1(j), which interprets the appearance of sharply deformed hexagonal H domains in the right part of Fig. 1(a).

Both our experimental and theoretical results above demonstrate that structural reconstruction induced by vdW interlayer interaction and external heterostrain greatly influence the geometry of the long-wavelength moiré superlattice, therefore, are expected to have significant impact on its electronic properties. We first study the electronic properties of the reconstructed WS$_2$ moiré superlattice with small heterostrain in Fig. 1(b). Fig. 2(a) shows the typical *dI/dV* spectra taken at H, DW, XX′ and MM′ stackings, respectively (see Fig. S2 for more STS spectra). The indirect band gap is about 1eV in the *dI/dV* spectra which is consistent with the one reported in multilayer WS$_2$[45,46]. Characteristic features of these *dI/dV* spectra, such as the locations of the critical points $K_V$, $\Gamma_V$ in valence band and $Q_C$, $K_C$ in conduction band, can be identified through comparing with the first-principles calculated band structure and layer-resolved local density of state (LDOS) of H-stacked multilayer WS$_2$ in Fig. 2(b). We notice that the width of the characteristic peaks in our spectra is large because of the relatively high temperature of our measurement. Our main finding from the STS measurement is that the location of $K_C$ point in conduction band obviously varies at different stacked regions, as summarized in Fig. 2(c), in contrast to the locations of other critical points that are almost unchanged.

Generally, the band structure of a commensurate vdW materials can be affected by stacking order[47], interlayer spacing[23,47] and strain-induced deformation[23,48-52]. In order to find out the main contributor in our experiment, we carry out first-principles calculation considering separately these three factors. First, we calculate band structures and LDOS of DW, XX′ and MM′-stacked multilayer $WS_2$ with fixed lattice constant and interlayer spacing in Fig. S3. The location of $K_C$ point in XX′ stacking slightly shifts towards higher energy, while the ones in other stackings are almost same, which disagrees with our experimental result in Fig. 2(a). Second, we calculate the LDOS of a DW-stacked multilayer $WS_2$ with various interlayer spacings ranging from 6.00Å to 6.77Å in Fig. S4. We find that, although the locations of $K_V$, $\Gamma_V$ in valence band dramatically shift with the change of interlayer spacing, the locations of $Q_C$, $K_C$ point show slight variations which are hard to be detected in our measurement resolution.

Having excluded the former two factors as the essential reasons to induce the electronic band modulation, we now turn to study the effect of strain-induced lattice deformation which is contributed from both vdW interlayer coupling and external intralayer heterostrain in our experiment. The following analysis of lattice deformation and band modulation in different stacked regions are both taken relative to that in commensurate H-stacked region. Through measuring the distances between adjacent S atoms along zigzag direction ($l_z$) and along armchair direction ($l_a$) in the atomic-resolved STM images, we could summarize the strain-induced lattice deformation ratio $s$ (%) of the four different stacked region relative to H-stacked region at zigzag direction (blue dots, $s_z$) and armchair direction (red dots, $s_a$), respectively, as shown in Fig. 2(d) (see Fig. S5 for details of lattice deformation analysis). The lattices in XX′-stacked region show an obvious tensile deformation at zigzag direction, while the lattices in MM′-stacked region mainly have a compressive deformation at armchair direction. Considering these lattice deformation data, Fig. 2(e) shows the calculated LDOS of different stackings (see Fig. S6 for the corresponding band structures). We find that $K_C$ point shifts towards higher energy in MM′-stacked region and shifts towards lower energy in both DW and XX′-stacked region, in contrast to the locations of $K_V$, $\Gamma_V$ points that are almost unchanged, which is accorded with our experimental result in Fig. 2(a). We note that the energy shift of $K_C$ point in the theoretical calculation is smaller than that detected in our experiment, which may be resulted from the error of the detected lattice deformation ratio in Fig. 2(d).

The experimental and theoretical results in Fig. 2 demonstrate that the strain-induced deformation greatly influences the location of $K_C$ point of conduction band in the long-wavelength reconstructed $WS_2$ moiré superlattice. In order to quantify the dependence between the band modulation and the strain-induced deformation, we carry out STS measurement on the sharply deformed moiré region with stronger external intralayer heterostrain. According to the result in Fig. 1 and Fig. S1, the 1D DWs in the deformed moiré superlattice, with the length ranging from 10 nm to hundreds of nanometers tuned by external heterostrain, provides a great platform to study strain-engineered effect[48]. Fig. 3(a) shows the STM image of a typical sharply deformed moiré superlattice. One of the DWs, named as $DW_1$, presents much larger height in the STM topography, with the length of about 45nm and the width of about 18nm. An atomic-resolved STM image in Fig. 3(b) shows the existence of an intersection angle 11° between the direction of $DW_1$ and the zigzag direction. Detailed analysis of lattice deformation across the $DW_1$ performed along armchair direction and zigzag direction are given in Fig. 3(d) and Fig. S7, respectively. The compressive lattice deformation is obviously detected at armchair direction across $DW_1$ and shows larger strength when closing to the central position of $DW_1$, which is well fitted as a Lorentzian function of positions in Fig. 3(d).

Fig. 3(e) shows the $dI/dV$ spectrum taken at H-stacked region of this deformed moiré superlattice, in which the locations of $Q_C$ and $K_C$ point in conduction band are identified at 0.20eV and 0.25eV. These locations of the critical points are different from that detected in Fig. 2(a), because the doping induced by charge transfers of underlying layers is different in these two regions. We then take $dI/dV$ spectra map as a function of positions across $DW_1$ in Fig. 3(f), showing the shift of $K_C$ point over a smaller energy range. Here, the substrate baseline of the $dI/dV$ spectra in Fig. 3(f) is deducted in order to make the characteristic peaks clearer (see Fig. S8). Remarkably, the peak position of $K_C$ point in conduction band moves towards higher bias voltage when coming close to the center of $DW_1$, while the peak position of $Q_C$ point keeps almost unchanged across the whole $DW_1$. The energy shift $\Delta E$ of $K_C$ point is summarized in Fig. 3(g) with its maximum reaching as high as 300meV, which is well fitted as a Lorentzian function of positions across the $DW_1$.

The above experimental observations are well reproduced by our first-principles calculations in Fig. 4(a) and Fig. S9. The location of $K_C$ point in calculated LDOS moves noticeably towards higher energy when adding a 1.78% compressive lattice deformation at armchair direction in DW, while the location of $Q_C$ point does not show obvious shift. Taking the experimental lattice deformation ratio at armchair direction in Fig. 3(d), Fig. 4(b) shows the calculated LDOS map as a function of positions across $DW_1$ with the substrate baseline deducted (see Fig. S9). In agree well with our experiment result in Fig. 3(f), the calculated energy shift of $K_C$ point increases monotonously with increasing the strength of lattice deformation, and reaches its maximum at the central position of $DW_1$.

The sensitive strain-dependent electronic modulation in turn presents as an efficient way to infer the lattice deformation in the reconstructed moiré superlattice. Fig. 4(c) shows the experimental energy map for the energy shift $\Delta E$ of $K_C$ point as a function of positions detected in the area around $DW_1$. We find that the strain-induced electronic modulation is observed not only in the direction across the $DW_1$, but also in the direction along $DW_1$, which shows that the energy shift $\Delta E$ of $K_C$ point becomes larger when getting closer to the XX′-stacked region along $DW_1$ (see Fig. S10 for the measurements along $DW_1$). This result reflects the lattice deformation information that the reconstruction-induced strain along the 1D $DW_1$ is inhomogeneity and becomes larger near XX′-stacked region.

In summary, we have performed a combined experimental and theoretical study on strain-dependent structural reconstruction and electronic structure modulation at nanometer scale in long-wavelength $WS_2$ moiré superlattice. The external intralayer heterostrain tunes greatly both the geometry and the moiré wavelength of the reconstructed moiré superlattice. Importantly, the $K$ point in the conduction band is sensitively modulated by strain-induced lattice deformation in this system, with the maximum modulation reaching over 300meV. Our results highlight that strain effect plays an essential role on determining accurate structural the electronic properties of TMD moiré superlattice.

**Acknowledgements:**

This work was supported by the National Natural Science Foundation of China (Grant Nos. 12104144, 62090035, U19A2090 and 11904095); the National Natural Science Foundation of Hunan Province, China (Grant No. 2021JJ20025); the Key Program of Science and Technology Department of Hunan Province (2019XK2001, 2020XK2001); National Key Research and Development Program of Ministry of Science and Technology (2021YFA1200503).

Figures:

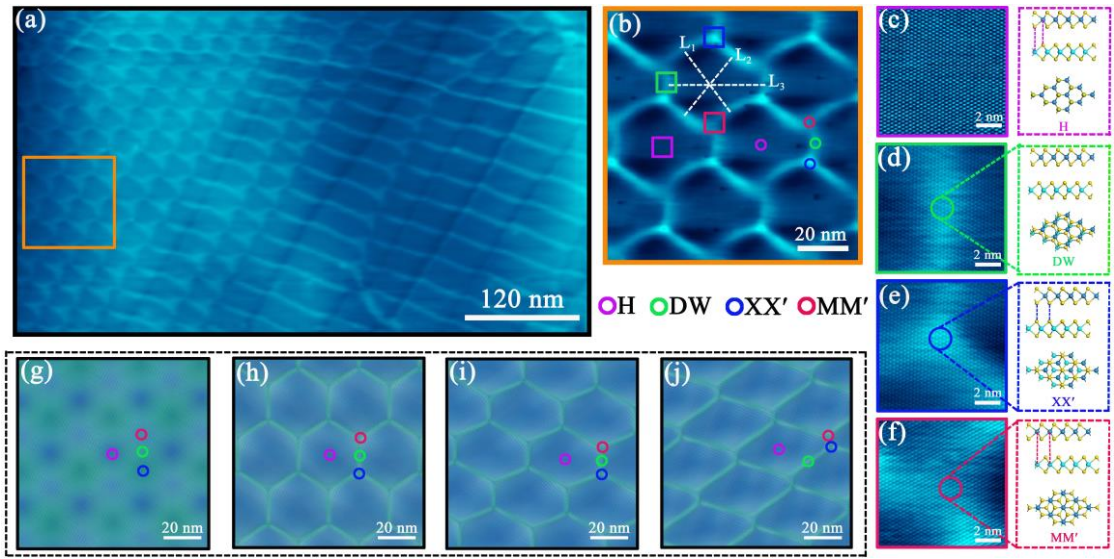

**FIG.1 (a)** A 700nm×400nm STM image for long-wavelength WS$_2$ moiré superlattices ($V_{sample}$ = 1.5V and $I$ = 100pA). **(b)** A 100nm×100nm STM image of the reconstructed moiré superlattice containing hexagonal H-stacked domains separated by a dislocation network. The colored circles mark the corresponding stacking configurations of different locations. **(c)-(f)** 10nm×10nm atomic-resolved STM images taken around the H, DW, XX′ and MM′-stacked regions in panel (b), respectively. Their corresponding stacking configurations are schematically shown besides the STM images. **(g)-(j)** First-principles calculated structural evolution of the moiré superlattice in marginally twisted bilayer WS$_2$ ($\theta$ = 0.57°) considering lattice reconstruction [panel (h)] and external intralayer heterostrain [panel (i) and (j)].

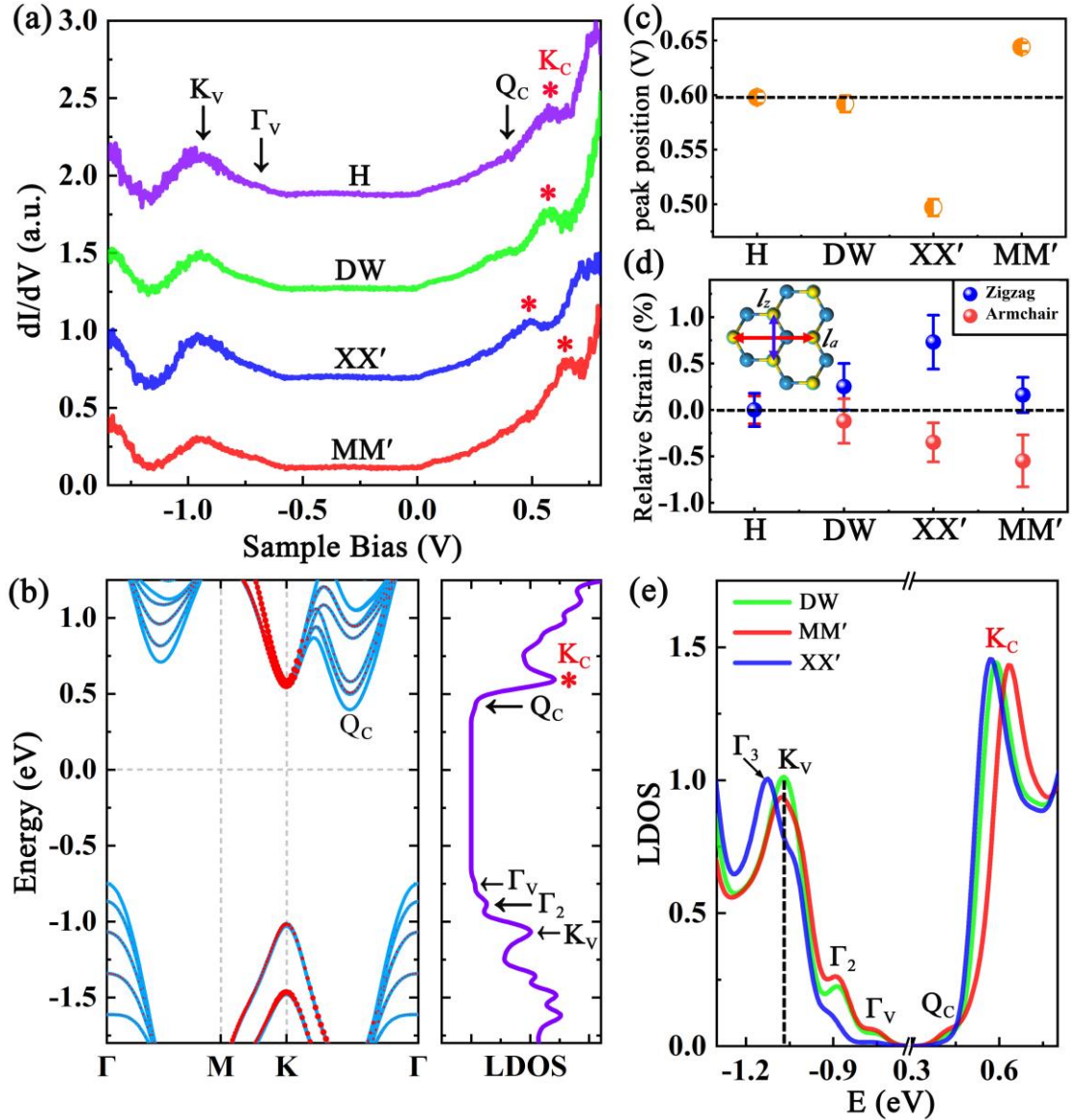

**FIG. 2 (a)** *dI/dV* spectra taken at H, DW, XX′ and MM′-stacked regions of the $WS_2$ moiré superlattice, respectively. **(b)** First-principles calculated band structure and corresponding LDOS of H-stacked multilayer $WS_2$ with the band extrema labelled. The color measures the weight projected to the topmost layer. **(c)** Summarized peak positions of $K_C$ point in *dI/dV* spectra of the four different stackings. **(d)** Summarized lattice deformation ratio *s* (%) of the four different stacked regions relative to H-stacked region at zigzag direction (blue dots) and armchair direction (red dots), respectively. Inset: $l_z$ and $l_a$ sign the distances of neighbour S atoms along zigzag and armchair directions, respectively. **(e)** Calculated LDOS for DW, XX′ and MM′-stacked multilayer $WS_2$ considering the same lattice deformation as the experimental results in panel (d).

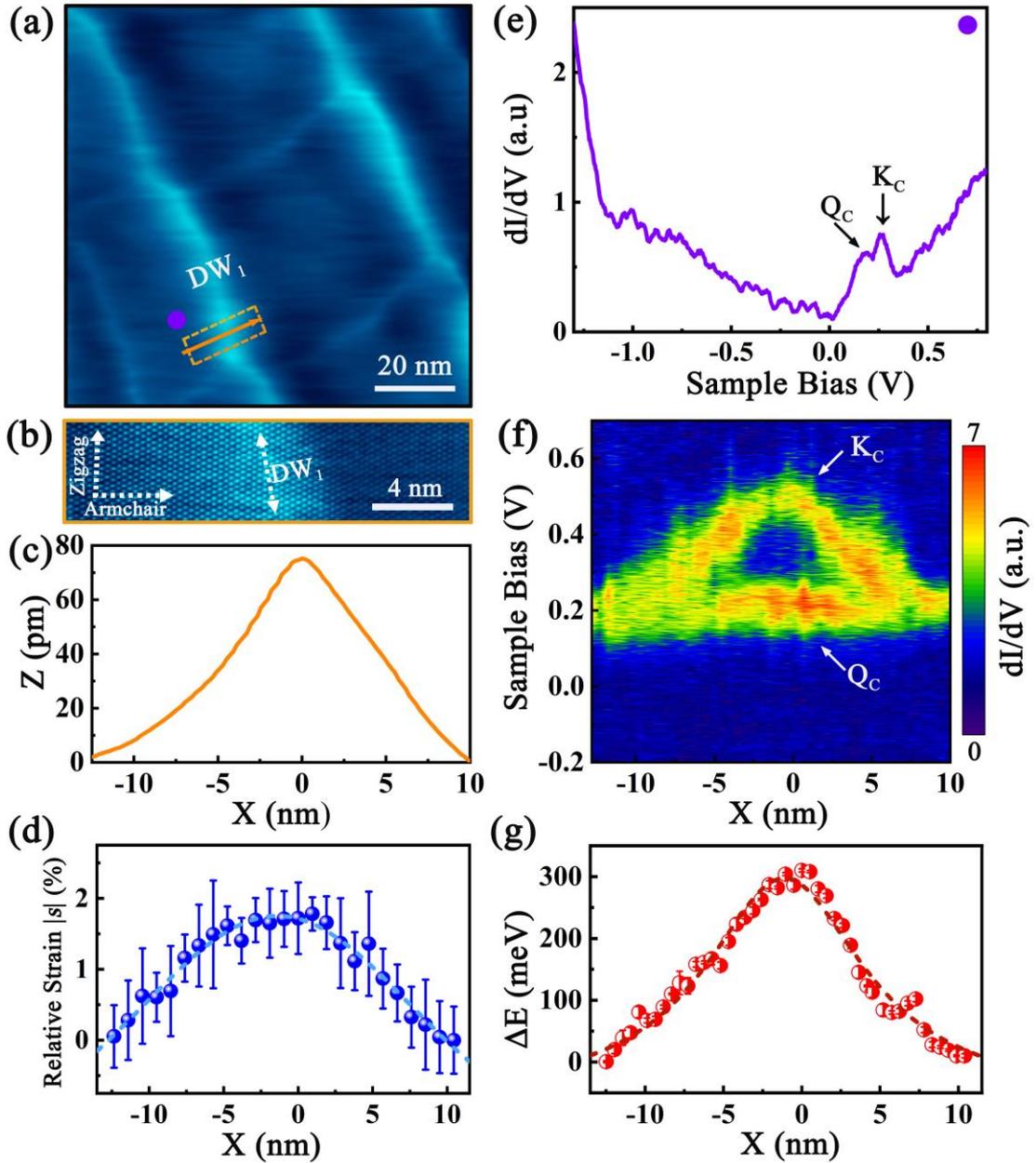

**FIG. 3 (a)** A 100nm×100nm STM image of a sharply deformed $WS_2$ moiré superlattice ($V_{sample}$ = 1.3V and $I$ = 100pA). **(b)** A 20nm×4nm atomic-resolved STM image across the $DW_1$ taken from the orange box in panel (a). **(c)** The profile line across the $DW_1$ taken along the orange arrow in panel (a). X = 0 is the central position of $DW_1$. **(d)** Summarized relative lattice deformation ratio $s_a$ (%) at armchair direction as a function of positions across the $DW_1$ along the profile line in panel (c). **(e)** $dI/dV$ spectrum of H-stacked region taken at the position marked by the purple dot in panel (a). **(f)** $dI/dV$ spectra map as a function of positions recorded across the $DW_1$ along the profile line in panel (c). **(g)** Summarized energy shift $\Delta E$ of $K_C$ point as a function of positions across the $DW_1$. The dashed lines in (d), (g) show the Lorentzian fitting curves.

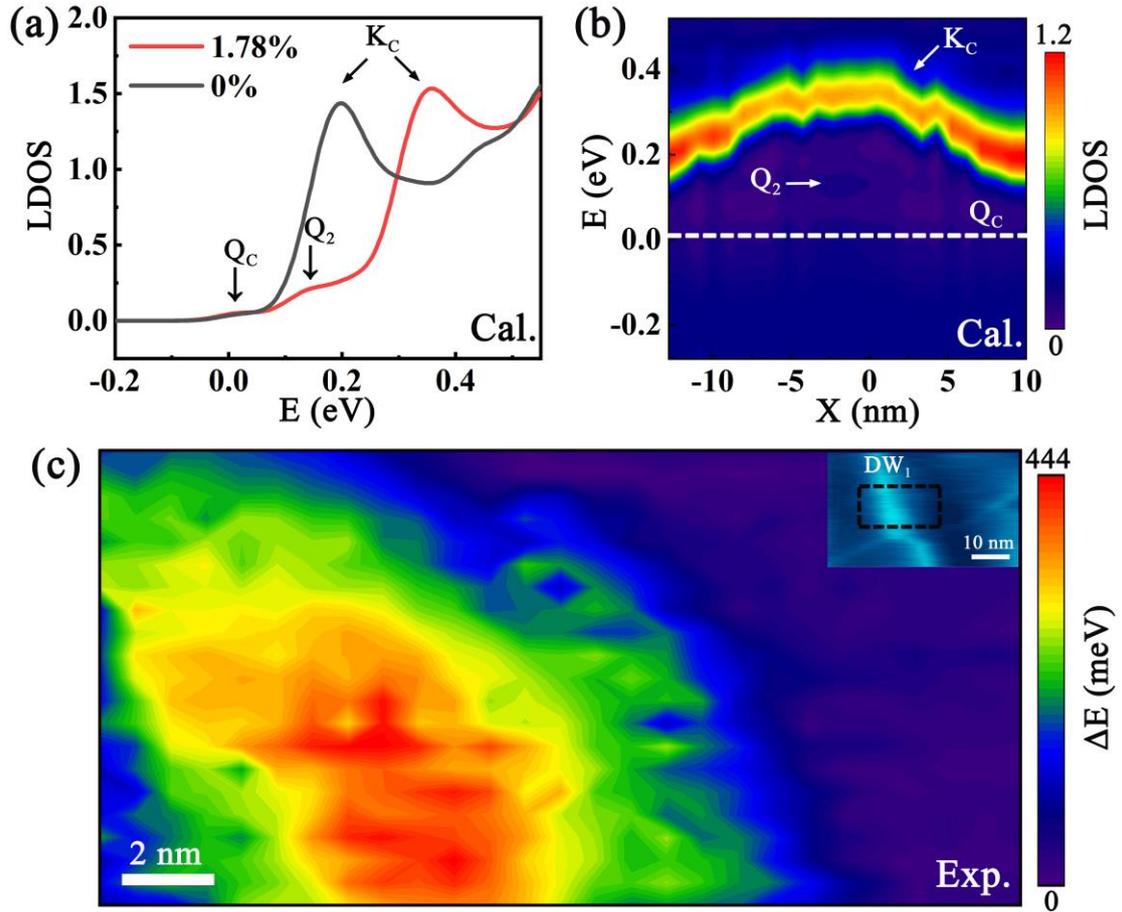

**FIG. 4 (a)** Calculated LDOS of DW-stacked multilayer WS$_2$ near the $K_C$ point in conduction band with 0% and 1.78% compressive lattice deformation along armchair direction, respectively. **(b)** Calculated LDOS map as a function of positions across DW$_1$. The substrate baseline of each LDOS is deducted. **(c)** The experimental energy map for the energy shift $\Delta E$ of $K_C$ point as a function of positions around DW$_1$, taken in the area marked by the black dashed square in the insert STM image.